\documentstyle[aps]{revtex}
\draft

\title{Quantum Networks for Elementary Arithmetic Operations}
\author{Vlatko Vedral\thanks{Current address: Blackett Laboratory,
    Imperial College, Prince Consort Road, London SW7 2BZ, U.K.},
  Adriano Barenco and Artur Ekert\\ {\protect\small\em Clarendon
    Laboratory, Department of Physics} \\ {\protect\small\em
    University of Oxford, Oxford, OX1 3PU, U.K.}}

\date{Submitted to Phys. Rev. A}

\begin{document}
\maketitle

\begin{abstract}
  Quantum computers require quantum arithmetic. We provide an explicit
  construction of quantum networks effecting basic arithmetic
  operations: from addition to modular exponentiation. Quantum modular
  exponentiation seems to be the most difficult (time and space
  consuming) part of Shor's quantum factorising algorithm. We show
  that the auxiliary memory required to perform this operation in a
  reversible way grows linearly with the size of the number to be
  factorised.
\end{abstract}


\section{Introduction}

A quantum computer is a physical machine that can accept input states
which represent a coherent superposition of many different possible
inputs and subsequently evolve them into a corresponding superposition
of outputs. Computation, {\em i.e.\/} a sequence of unitary
transformations, affects simultaneously each element of the
superposition, generating a massive parallel data processing albeit
within one piece of quantum hardware~\cite{DD85}.  This way quantum
computers can efficiently solve some problems which are believed to be
intractable on any classical computer~\cite{DJ92,PWS94}. Apart from
changing the complexity classes, the quantum theory of computation
reveals the fundamental connections between the laws of physics and
the nature of computation and mathematics~\cite{DDBook}.

For the purpose of this paper a quantum computer will be viewed as a
quantum network (or a family of quantum networks) composed of quantum
logic gates; each gate performing an elementary unitary operation on
one, two or more two--state quantum systems called {\em
  qubits\/}~\cite{DD89}. Each qubit represents an elementary unit of
information; it has a chosen ``computational'' basis
$\{|0\rangle,|1\rangle\}$ corresponding to the classical bit values
$0$ and $1$.  Boolean operations which map sequences of 0's and 1's
into another sequences of 0's and 1's are defined with respect to this
computational basis.

Any unitary operation is reversible that is why quantum networks
effecting elementary arithmetic operations such as addition,
multiplication and exponentiation cannot be directly deduced from
their classical Boolean counterparts (classical logic gates such as
{\small\sf AND} or {\small\sf OR} are clearly irreversible: reading
$1$ at the output of the {\small \sf OR} gate does not provide enough
information to determine the input which could be either $(0,1)$ or
$(1,0)$ or $(1,1)$). Quantum arithmetic must be built from
reversible logical components. It has been shown that reversible
networks (a prerequisite for quantum computation) require some
additional memory for storing intermediate results~\cite{RL61,CHB89}.
Hence the art of building quantum networks is often reduced to
minimising this auxiliary memory or to optimising the trade--off
between the auxiliary memory and a number of computational steps
required to complete a given operation in a reversible way.

In this paper we provide an explicit construction of several
elementary quantum networks. We focus on the space complexity {\em
  i.e.\/} on the optimal use of the auxiliary memory.
In our constructions, we save memory by reversing some computations
with different computations (rather than with the same computation but
run backwards~\cite{CHB89}). The networks are presented in the
ascending order of complication. We start from a simple quantum
addition, and end up with a modular exponentiation
\begin{equation}
 U_{a,N}|x\rangle\otimes|0\rangle
 \rightarrow
 |x\rangle\otimes|a^x \bmod N\rangle,
\label{unitary}
\end{equation}
where $a$ and $N$ are predetermined and known parameters.
This particular operation plays an important role in Shor's quantum
factoring algorithm~\cite{PWS94} and seems to be its most demanding
part.

The structure of the paper is as follows: in Section~\ref{tools} we
define some basic terms and describe methods of reversing some types
computation, in Section~\ref{networks} we provide a detailed
description of the selected quantum networks and in
Section~\ref{complex} we discuss their complexity.

\section{Basic concepts}
\label{tools}

For completeness let us start with some basic definitions. A quantum
network is a quantum computing device consisting of quantum logic
gates whose computational steps are synchronised in time. The outputs
of some of the gates are connected by wires to the inputs of others.
The size of the network is its number of gates. The size of the input
of the network is its number of input qubits {\em i.e.\/} the qubits
that are prepared appropriately at the beginning of each computation
performed by the network. Inputs are encoded in binary form in the
computational basis of selected qubits often called a {\em quantum
  register\/}, or simply a {\em register\/}. For instance, the binary
form of number $6$ is $110$ and loading a quantum register with this
value is done by preparing three qubits in state $|1\rangle \otimes
|1\rangle \otimes |0\rangle$.  In the following we use a more compact
notation: $|a\rangle$ stands for the direct product $|a_n\rangle
\otimes |a_{n-1}\rangle \ldots|a_1\rangle \otimes|a_0\rangle$ which
denotes a quantum register prepared with the value $a=2^0 a_0+ 2^1 a_1
+\ldots 2^n a_n$. Computation is defined as a unitary evolution of the
network which takes its initial state ``input'' into some final state
``output''.

Both the input and the output can be encoded in several registers.
Even when $f$ is a one--to--one map between the input $x$ and the
output $f(x)$ and the operation can be formally written as a unitary
operator $U_f$
\begin{equation}
  U_f|x\rangle\rightarrow |f(x)\rangle,
\label{bij}
\end{equation}
we may still need an auxiliary register to store the intermediate
data. When $f$ is not a bijection we have to use an additional
register in order to guarantee the unitarity of computation. In this
case the computation must be viewed as a unitary transformation $U_f$
of (at least) two registers
\begin{equation}
  U_f |x,0\rangle \rightarrow |x,f(x)\rangle,
\end{equation}
where the second register is of appropriate size to accommodate
$f(x)$.

As an example, consider a function $f_{a,N}: x\rightarrow ax\bmod N$.
A quantum network that effects this computation takes the value $x$
from a register and multiplies it by a parameter $a$ modulo another
parameter $N$. If $a$ and $N$ are coprime, the function is bijective
in the interval $\{ 0,1, \ldots, N-1 \}$, and it is possible to
construct a network that writes the answer into the same register
which initially contained the input $x$ (as in the equation
(\ref{bij})).  This can be achieved by introducing an auxiliary
register and performing
\begin{equation}
  U_{a,N}|x,0\rangle \rightarrow |x,ax\bmod N\rangle.
\end{equation}
Then we can precompute $a^{-1}\bmod N$, the inverse of $a$ modulo $N$
(this can be done classically in an efficient way using Euclid's
algorithm~\cite{DEK81}), and, by exchanging the two registers and
applying $U^{-1}_{a^{-1}\bmod N,N}$ to the resulting state, we obtain
\begin{equation}
  U^{-1}_{a^{-1}\bmod N,N} S|x, ax\bmod N\rangle \rightarrow
  U^{-1}_{a^{-1}\bmod N,N} |ax\bmod N,x\rangle \rightarrow |ax\bmod
    N,0\rangle,
\end{equation}
where $S$ is a unitary operation that exchanges the states of the two
registers. Thus,
\begin{equation}
  U^{-1}_{a^{-1}\bmod N,N} SU_{a,N}|x,0\rangle \rightarrow
| ax\bmod N,0\rangle
\end{equation}
effectively performs
\begin{equation}
  |x\rangle \rightarrow |f(x)\rangle
\end{equation}
where the second register is treated as an internal part of the
network (temporary register).

\section{Network architecture}
\label{networks}

Quantum networks for basic arithmetic operations can be constructed in
a number of different ways. Although almost any non-trivial quantum
gate operating on two or more qubits can be used as an elementary
building block of the networks~\cite{AB95} we have decided to use the
three gates described in Fig.~\ref{basicgates}, hereafter refered to
as {\em elementary gates}. None of these gates is universal for
quantum computation, however, they suffice to build any Boolean
functions as the Toffoli gate alone suffices to support any {\em
  classical} reversible computation. The {\small\sf NOT} and the
Control--{\small \sf NOT} gates are added for convenience (they can be
easily obtained from the Toffoli gates).

\subsection{Plain adder}

The addition of two registers $|a\rangle$ and $|b\rangle$ is probably
the most basic operation, in the simplest form it can be written as
\begin{equation}
  |a,b,0\rangle \rightarrow |a,b,a+b\rangle.
\end{equation}
Here we will focus on a slightly more complicated (but more useful)
operation that rewrites the result of the computation into the one of
the input registers , {\em i.e.\/}
\begin{equation}
  |a,b\rangle \rightarrow |a,a+b\rangle,
\end{equation}
As one can reconstruct the input $(a,b)$ out of the output $(a,a+b)$,
there is no loss of information, and the calculation can be
implemented reversibly. To prevent overflows, the second register
(initially loaded in state $|b\rangle$) should be sufficiently large,
{\em i.e.\/} if both $a$ and $b$ are encoded on $n$ qubits, the second
register should be of size $n+1$. In addition, the network described
here also requires a temporary register of size $n-1$, initially in
state $|0\rangle$, to which the carries of the addition are
provisionally written (the last carry is the most significant bit of
the result and is written in the last qubit of the second register).

\begin{figure}
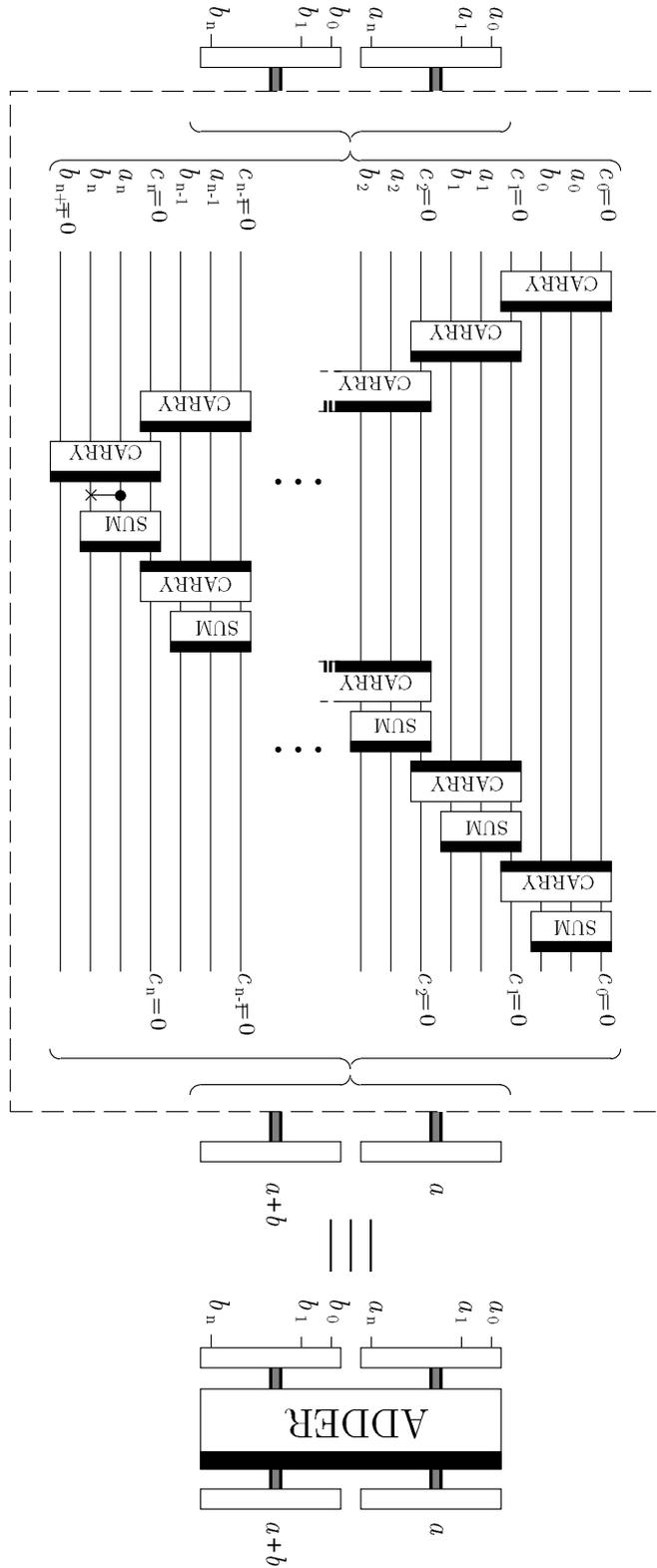

\vspace{4mm}
\centerline{
}
\vspace{2mm}
\caption[fo2]{ \small
  Plain adder network. In a first step, all the carries are calculated
  until the last carry gives the most significant digit of the result.
  Then all these operations apart from the last one are undone in
  reverse order, and the sum of the digits is performed
  correspondingly. Note the position of a thick black bar on the right
  or left hand side of basic carry and sum networks.  A network with a
  bar on the left side represents the reversed sequence of elementary
  gates embeded in the same network with the bar on the right side.  }
\label{plainadder}
\end{figure}

The operation of the full addition network is illustrated in
Fig.~\ref{plainadder} and can be understood as follows:
\begin{itemize}
\item
We compute the most significant bit of the result $a+b$. This step
requires computing all the carries $c_i$ through the relation $c_i
\leftarrow a_i$ {\small \sf AND} $b_i$ {\small \sf AND} $c_{i-1}$,
where $a_i$, $b_i$ and $c_i$ represent the $i$th qubit of the first,
second and temporary (carry) register respectively. Fig.~\ref{carrysum}i)
illustrates the sub--network that effects the carry calculation.
\item Subsequently we reverse all these operations (except for the
  last one which computed the leading bit of the result) in order to
  restore every qubit of the temporary register to its initial state
  $|0\rangle$. This enables us to reuse the same temporary register,
  should the problem, for example, require repeated additions. During
  the resetting process the other $n$ qubits of the result are
  computed through the relation $b_i \leftarrow a_i$ {\small \sf XOR}
  $b_i$ {\small \sf XOR} $c_{i-1}$ and stored in the second register.
  This operation effectively computes the $n$ first digits of the sum
  (the basic network that performs the summation of three qubits
  modulo $2$ is depicted in Fig.~\ref{carrysum}ii).)
\end{itemize}

If we reverse the action of the above network ({\em i.e.\/} if we
apply each gate of the network in the reversed order) with the input
$(a,b)$, the output will produce $(a,a-b)$ when $a\geq b$. When $a<b$,
the output is $(a,2^{n+1}-(b-a))$, where $n+1$ is the size of the
second register. In this case the most significant qubit of the second
register will always contain $1$ . By checking this ``overflow bit''
it is therefore possible to compare the two numbers $a$ and $b$; we
will use this operation in the network for modular addition.

\subsection{Adder modulo $N$}

A slight complication occurs when one attempts to build a network that
effects
\begin{equation}
  |a,b\rangle \rightarrow |a,a+b \bmod N\rangle,
\end{equation}
where $0\leq a,b <N$.  As in the case of the plain adder, there is no
a priori violation of unitarity since the input $(a,b)$ can be
reconstructed from the output $(a,a+b \bmod N)$, when $0\leq a,b<N$
(as it will always be the case).  Our approach is based on taking the
output of the plain adder network, and subtracting $N$, depending on
whether the value $a+b$ is bigger or smaller than $N$.  The method,
however, must also accomodate a superposition of states for which some
values $a+b$ are bigger than $N$ and some smaller than $N$.

Fig.~\ref{addmodn} illustrates the various steps needed to implement
modular addition. The first adder performs a plain addition on the
state $|a,b\rangle$ returning $|a,a+b\rangle$; the first register is
then swapped with a temporary register formerly loaded with $N$, and a
subtractor ({\em i.e.\/} an adder whose network is run backwards) is
used to obtain the state $|N,a+b-N\rangle$.  At this stage the most
significant bit of the second register indicates whether or not an
overflow occurred in the subtraction, {\em i.e.\/} whether $a+b$ is
smaller than $N$ or not. This information is ``copied'' into a
temporary qubit $|t\rangle$ (initially prepared in state $|0\rangle$)
through the Control--{\small \sf NOT} gate.  Conditionally on the
value of this last qubit $|t\rangle$, $N$ is added back to the second
register, leaving it with the value $a+b \bmod N$. This is done by
either leaving the first register with the value $N$ (in case of
overflow), or resetting it to $0$ (if there is no overflow) and then
using a plain adder. After this operation, the value of the first
register can be reset to its original value and the first and the
temporary register can be swapped back, leaving the first two
registers in state $|a, a+b \bmod N\rangle$ and the temporary one in
state $|0\rangle$. At this point the modular addition has been
computed, but some information is left in the temporary qubit
$|t\rangle$ that recorded the overflow of the subtraction. This
temporary qubit cannot be reused in a subsequent modular addition,
unless it is coherently reset to zero. The last two blocks of the
network take care of this resetting: first the value in the first
register ($=a$) is subtracted from the value in the second ($=a+b
\bmod N$) yielding a total state $|a,(a+b \bmod N)-a\rangle$. As
before, the most significant bit of the second register contains the
information about the overflow in the subtraction, indicating whether
or not the value $N$ was subtracted after the third network. This bit
is then used to reset the temporary bit $|t\rangle$ to $|0\rangle$
through a second Control--{\small \sf NOT} gate. Finally the last
subtraction is undone, returning the two registers to the state
$|a,a+b \bmod N\rangle$.

\subsection{Controlled--multiplier modulo $N$}

Function $f_{a,N}(x)=ax \bmod N$ can be implemented by repeated
conditional additions (modulo $N$): $ax=2^0 a x_0+2^1 a x_1 + \ldots
2^{n-1} a x_{n-1}$. Starting from a register initially in the state
$|0\rangle$, the network consists simply of $n$ stages in which the
value $2^i a$ is added conditionally, depending on the state of the
qubit $|x_i\rangle$. Fig.~\ref{controlmult} shows the corresponding
network; it is slightly complicated by the fact that we want the
multiplication to be effected conditionally upon the value of some
external qubit $|c\rangle$, namely, we want to implement
\begin{equation}
 |c;x,0\rangle\rightarrow \left\{
\begin{array}{ll}
 |c;x,a\times x\bmod N\rangle & \mbox{if $|c\rangle=|1\rangle$}
\\ \nonumber
 |c;x,x\rangle & \mbox{if $|c\rangle=|0\rangle$}
\end{array}
\right.
\end{equation}
To account for this fact at the $i$th modular addition stage the first
register is loaded with the value $2^i a$ if
$|c,x_i\rangle=|1,1\rangle$ and with value $0$ otherwise.
This is done by applying the Toffoli gate to the control qubits
$|c\rangle$ and $|x_i\rangle$ and the appropriate target
qubit in the register; the gate is applied each time value ``$1$''
appears in the binary form of the number $2^ia$.

Resetting the register to its initial state is done by applying the
same sequence of the Toffoli gates again (the order of the gates is
irrelevant as they act on different target qubits).  If
$|c\rangle=|0\rangle$ only $0$ values are added at each of the $n$
stages to the result register giving state $|c;x,0\rangle$.  Since we
want the state to be $|c;x,x\rangle$ we copy the content of the input
register to the result register if $|c\rangle=|0\rangle$. This last
operation is performed by the rightmost elements of the network of
Fig.~\ref{controlmult}.  The conditional copy is implemented using an
array of Toffoli gates.

\subsection{Exponentiation Modulo N}

A reversible network that computes the function $f_{a,N}(x)=a^x \bmod
N$ can now be designed using the previous constructions. Notice first
that $a^x$ can be written as $a^x=a^{2^0x_0}\cdot a^{2^1 x_1} \cdot
\ldots a^{2^{m-1} x_{m-1}}$, thus modular exponentiation can be
computed by setting initially the result register to $|1\rangle$, and
successively effecting $n$ multiplications by $a^{2^i}$ (modulo $N$)
depending on the value of the qubit $|x_i\rangle$; if $x_i=1$, we want
the operation
\begin{equation}
|a^{2^0x_0+\ldots 2^{i-1}x_{i-1}},0\rangle \rightarrow
|a^{2^0x_0+\ldots 2^{i-1}x_{i-1}},a^{2^0x_0+\ldots
    2^{i-1}x_{i-1}}\cdot a^{2^i}\rangle
\end{equation}
to be performed, otherwise, when
$x_i=0$ we just require
\begin{equation}
|a^{2^0x_0+\ldots 2^{i-1}x_{i-1}},0\rangle
\rightarrow |a^{2^0x_0+\ldots 2^{i-1}x_{i-1}},a^{2^0x_0+\ldots
    2^{i-1}x_{i-1}}\rangle.
\end{equation}
Note that in both cases the result can be written as
$|a^{2^0x_0+\ldots 2^{i-1}x_{i-1}},a^{2^0x_0+\ldots 2^ix_i}\rangle$.
To avoid an accumulation of intermediate data in the memory of the
quantum computer, a particular care should be taken to erase the
partial information generated. This is done, as explained in
Sect.~\ref{tools}, by running backwards a controlled multiplication
network with the value $a^{-2^i} \bmod N$. This quantity can be
efficiently precomputed in a classical way~\cite{DEK81}.
Fig.~\ref{modexp} shows the network for a complete modular
exponentiation. It is made out of $m$ stages; each stage performs the
following sequence of operations:
\begin{equation}
\begin{array}{ll}
 |a^{2^0x_0+\ldots 2^{i-1}x_{i-1}},0\rangle
\rightarrow &\hspace{2cm} \mbox{\small (multiplication)}\\ \nonumber
\hspace{2cm}
 |a^{2^0x_0+\ldots 2^{i-1}x_{i-1}}, a^{2^0x_0+\ldots 2^ix_i}\rangle
\rightarrow &\hspace{2cm}  \mbox{\small (swapping)}\\ \nonumber
\hspace{2cm}
| a^{2^0x_0+\ldots 2^ix_i},a^{2^0x_0+\ldots 2^{i-1}x_{i-1}}\rangle
\rightarrow &\hspace{2cm}  \mbox{\small (resetting)}\\ \nonumber
\hspace{2cm}
| a^{2^0x_0+\ldots 2^ix_i},0\rangle
\end{array}
\end{equation}

\section{Network complexity}
\label{complex}

The size of the described networks depends on the size of their input
$n$. The number of elementary gates in the plain adder, the modular
addition and the controlled--modular addition network scales linearly
with $n$. The controlled modular multiplication contains $n$
controlled modular additions, and thus requires of the order of $n^2$
elementary operations.  Similarly the network for exponentiation
contains of the order of $n$ controlled modular multiplications and
the total number of elementary operations is of the order of $n^3$.
The multiplicative overhead factor in front depends very much on what
is considered to be an elementary gate. For example, if we choose the
Control--{\small\sf NOT} to be our basic unit then the Toffoli gate
can be simulated by $6$ Control--{\small\sf NOT} gates ~\cite{ABC95}.

Let us have a closer look at the memory requirements for the modular
exponentiation; this can help to asses the difficulty of quantum
factorisation. We set $n$ to be the number of bits needed to encode
the parameter $N$ of Eq.~(\ref{unitary}). In Shor's algorithm, $x$ can
be as big as $N^2$, and therefore the register needed to encode it
requires up to $2 n$ qubits. Not counting the two input registers and
an additional bit to store the most significant digit of the result,
the plain adder network requires an extra $(n-1)$--qubit temporary
register for storing temporary (carry) qubits.  This register is reset
to its initial value, $|0\rangle$, after each operation of the network
and can be reused later.  The modular addition network, in addition to
the temporary qubit needed to store overflows in subtractions,
requires another $n$--qubit temporary register; in total this makes
two $n$--qubit temporary registers for modular addition.  Controlled
modular multiplication is done by repeated modular additions, and
requires three temporary $n$--qubit registers: one for its own
operation and two for the modular addition (controlled modular
multiplication also requires a temporary qubit used by the modular
addition network).  Finally, the network for exponentiation needs four
temporary $n$--qubit registers, one for its own operation and three
for the controlled modular multiplication (plus an additional qubit
used by the modular addition).  Altogether the total number of qubits
required to perform the first part of the factorisation algorithm is
$7n+1$, where $2n$ qubits are used to store $x$, $n$ qubits store the
result $a^x \bmod N$ and $4n+1$ qubits are used as temporary qubits.

The networks presented in this paper are by no means the only or the
most optimal ones. There are many ways to construct operation such as
$a^x\bmod N$, given parameters $a$ and $N$. Usually a dedicated
network composed of several sub--units does not have to be a simple
sum of the sub--units. In the modular exponentiation, for example, it
is relatively easy to reduce the memory {\em i.e.\/} the constant
overhead factor ($7$ in our case) by noting that the first register in
the plain adder network always stores  specific classical values:
either $0$ and $N$. The same holds for the temporary register in the
adder modulo $N$ which always stores either $0$ and $2^i a \bmod N$.
There is no need to use a full quantum register for this: a classical
register plus a single qubit (that keeps track  of the entanglement)
are sufficient. This reduces the number of qubits to $5n+2$. One
further register can be removed by using the addition network that
does not require a temporary register~\cite{GPZ95}; the trick is to
use the $n$--bit Toffoli gates to add $n$--bit numbers. If the
difficulty of the practical implementations of the $n$--bit Toffoli
gates is comparable to that of the regular Toffoli gate, then this can
be a good way of saving memory. All together the number of qubits can
be reduced from $7n+1$ to $4n+3$. This means that apart from the
register storing $x$ and another one storing $a^x \bmod N$ we need
additional $n+3$ temporary qubits to perform quantum modular
exponentiation in Shor's algorithm.  The required memory grows only as
a linear function of the size of $N$.

\section{Conclusion}

In this paper we have explicitly constructed quantum networks
performing elementary arithmetic operations including the modular
exponentiation which dominates the overall time and memory complexity
in Shor's quantum factorisation algorithm. Our network for the modular
exponentiation achieves only a linear growth of auxiliary memory by
exploiting the fact that $f_{a,N}(x)=ax \bmod N$ is a bijection (when
$a$ and $N$ are coprime) and can be made reversible by simple
auxiliary computations. In more practical terms our results indicate
that with the ``trapped ions computer''~\cite{CZ95} about $20$ ions
suffice (at least in principle) to factor $N=15$. Needless to say, the
form of the actual network that will be used in the first quantum
computer will greatly depend on the type of technology employed; the
notion of an optimal network is architecture dependent and any further
optimisation has to await future experimental progress.

\section{Acknowledgments}

V.~V. thanks the Royal Society for the vacation scholarship which
enabled him to undertake the research project on the subject of the
paper. A.~B. acknowledges the financial support of the Berrows
Fund at Lincoln College, Oxford.

The authors would like to thank D.~Deutsch, D.~DiVincenzo,
S.~Gardiner, H.J.~Kimble, P.L.~Knight, E.~Knill, T.~Pellizzari, and
P.~Zoller for useful discussions.

\begin{figure}
\vspace{4mm}
\centerline{
}
\vspace{2mm}
\caption[fo1]
{ \small Truth tables and graphical representations of the elementary
  quantum gates used for the construction of more complicated quantum
  networks. The control qubits are graphically represented by a dot,
  the target qubits by a cross. i) {\small\sf NOT} operation. ii)
  Control--{\small \sf NOT}.  This gate can be seen as a ``copy
  operation'' in the sense that a target qubit ($b$) initially in the
  state $0$ will be after the action of the gate in the same state as
  the control qubit. iii) Toffoli gate. This gate can also be seen as
  a Control--control--{\small \sf NOT}: the target bit ($c$) undergoes
  a {\small \sf NOT} operation only when the two controls ($a$ and
  $b$) are in state $1$. }
\label{basicgates}
\end{figure}

\begin{figure}
\vspace{4mm}
\centerline{
}
\vspace{2mm}
\caption[fo1b]{ \small
  Basic carry and sum operations for the plain addition network. i)
  the carry operation (note that the carry operation perturbs the
  state of the qubit $b$). ii) the sum operation.}
\label{carrysum}
\end{figure}

\begin{figure}
\vspace{4mm}
\centerline{
}
\caption[fo3]{ \small
  Adder modulo $N$. The first and the second network add $a$ and $b$
  together and then subtract $N$. The overflow is recorded into the
  temporary qubit $|t\rangle$. The next network calculates $(a+b)
  \bmod N$.  At this stage we have extra information about the value
  of the overflow stored in $|t\rangle$. The last two blocks restore
  $|t\rangle$ to $|0\rangle$.  The arrow before the third plain adder
  means that the first register is set to $|0\rangle$ if the value of
  the temporary qubit $|t\rangle$ is $1$ and is otherwise left
  unchanged (this can be easily done with Control--{\small \sf NOT}
  gates, as we know that the first register is in the state
  $|N\rangle$). The arrow after the third plain adder resets the first
  register to its original value (here $|N\rangle$). The significance
  of the thick black bars is explained in the caption of
  Fig.~\ref{plainadder}. }
\label{addmodn}
\end{figure}

\begin{figure}
\vspace{4mm}
\centerline{
}
\caption[fo5]{ \small
  Controlled multiplication modulo $N$ consists of consecutive modular
  additions of $2^ia$ or $0$ depending on the values of $c$ and $x_i$.
  The operation before the $i$th modular adder consists in storing
  $2^{i-1}a$ or $0$ in the temporary register depending on whether
  $|c,x_i\rangle=|1,1\rangle$ or not respectively.  Immediately after
  the addition has taken place, this operation is undone. At the end,
  we copy the content of the input register in the result register
  only if $|c\rangle=|0\rangle$, preparing to account for the fact
  that the final output state should be $|c;x,x\rangle$ and not
  $|c;x,0\rangle$ when $c=0$.  The signification of the thick black
  bars is given in the caption of Fig.~\ref{plainadder}.  }
\label{controlmult}
\end{figure}

\begin{figure}
\vspace{4mm}
\centerline{
}
\caption[fo6]{\small
  Modular exponentiation consists of successive modular
  multiplications by $a^{2^i}$. The even networks perform the reverse
  control modular multiplication by inverse of $a^{2^i} \bmod N$ thus
  resetting one of the registers to zero and freeing it for the next
  control modular multiplication. The signification of the thick black
  bars is given in the caption of Fig.~\ref{plainadder}.}
\label{modexp}
\end{figure}


\begin{thebibliography}{99}

\bibitem{DD85} D.~Deutsch, Proc. R. Soc. Lond. A  {\bf 400}, 97 (1985).

\bibitem{DJ92} D.~Deutsch and R.~Jozsa,  Proc. R. Soc. Lond. A {\bf
    439}, 553 (1992); E.~Bernstein and U.~Vazirani, in {\em Proc. 25th
    ACM Symposium on the Theory of Computation\/}, 11 (1993);
  D.S.~Simon, {\em Proceedings of the 35th Annual Symposium on the
    Foundations of Computer Science\/}, edited by S.~Goldwasser (IEEE
  Computer Society Press, Los Alamitos, CA), 16 (1994);

\bibitem{PWS94} P.W.~Shor, in {\em Proceedings of the 35th Annual
    Symposium on the Theory of Computer Science\/}, edited by S.
  Goldwasser (IEEE Computer Society Press, Los Alamitos, CA), p.124
  (1994).

\bibitem{DDBook} D.~Deutsch, {\em The Fabric of Reality\/}
  (Viking--Penguin Publishers, London, in print).

\bibitem{DD89} D.~Deutsch,  Proc. R. Soc. Lond. A {\bf 425}, 73 (1989).

\bibitem{RL61} R.~Landauer, IBM J. Res. Dev. {\bf 5}, 183 (1961);
  C.H~Bennett, IBM J. Res. Dev. {\bf 32}, 16 (1988); T.~Toffoli,
    Math. Systems Theory {\bf 14}, 13 (1981).

\bibitem{CHB89} C.H.~Bennett, SIAM J. Comput. {\bf 18(4)}, 766 (1989).

\bibitem{DEK81} D.E.~Knuth, {\em The Art of Computer Programming, Volume 2:
Seminumerical Algorithms\/} (Addison-Wesley, New York, 1981).

\bibitem{AB95} A.~Barenco, Proc. R. Soc. Lond. A, {\bf 449}, 679
  (1995); T.~Sleator and H.~Weinfurter, Phys. Rev. Lett. {\bf 74} 4087
  (1995); D.~Deutsch, A.~Barenco and A.~Ekert, Proc. R. Soc. Lond. A
  {\bf 449} 669 (1995); S.~Lloyd, Phys. Rev. Lett. {\bf 75}, 346
  (1995).

\bibitem{ABC95}  A.~Barenco, C.H.~Bennett, R.~Cleve, D.P.~DiVicenzo,
  N.~Margolus, P.~Shor, T.~Sleator, J.~Smolin and H.~Weinfurter,
  Phys.~Rev.~A {\bf 52}, 3457 (1995).

\bibitem{GPZ95} S.A.~Gardiner, T.~Pellizzari and P.~Zoller, {\em private
    communication}.

\bibitem{CZ95} J.I. Cirac and P. Zoller, Phys.~Rev.~Lett {\bf 74},
  4091 (1995).


\end{thebibliography}
\end{document}